\documentclass[letter, 12pt]{article}

\usepackage{amsmath}
\usepackage{graphicx, psfrag, epsf}
\usepackage{enumerate}
\usepackage{natbib}
\usepackage{url} 
\usepackage{multirow}

\usepackage[margin=1in]{geometry}

\usepackage{amssymb, amsfonts, amsthm, bm}
\usepackage{mathtools, mathrsfs,amsmath}
\usepackage{relsize} 
\usepackage{hyperref}
\usepackage{float}
\usepackage{enumitem}
\usepackage{enumerate}
\usepackage{algorithm,algpseudocode}
\usepackage{verbatim}

\newtheorem{assumption}{Assumption}

\DeclareMathOperator*{\argmax}{argmax} 

\DeclareMathOperator{\tr}{\mbox{tr}}

\usepackage{color}

\newcommand{\noop}[1]{}

\newcommand{\rrvert}{\vert}

\newcommand{\rrVert}{\Vert}
\newcommand{\llvert}{\vert}
\newcommand{\llVert}{\Vert}

\title{Prior distributions for Gaussian processes in computer model emulation and calibration}
\author{Mengyang Gu\thanks{
Department of Statistics and Applied Probability, University of California, Santa Barbara, CA, 93106, Email: \href{mailto:mengyang@pstat.ucsb.edu}{mengyang@pstat.ucsb.edu} }~ and Victor De Oliveira\thanks{Department of Management Science and Statistics, University of Texas at San Antonio, 
San Antonio, TX, 78249, Email: \href{mailto:victor.deoliveira@utsa.edu}{victor.deoliveira@utsa.edu } }}
\date{\today}

\begin{document}

\maketitle

\begin{abstract}

This article discusses prior distributions for the parameters of Gaussian processes (GPs) that are widely used as surrogate models to emulate expensive computer simulations. 
The parameters typically involve mean parameters, a variance parameter, and correlation parameters.  These parameters are often estimated by maximum likelihood (MLE). In some scenarios, however, the MLE can be unstable, particularly when the number of simulation runs is small, and some Bayesian estimators display better properties. 
We introduce default Bayesian priors for the parameters  of GPs with isotropic and separable correlation functions for emulating computer simulations 
with both scalar-valued and vector-valued outputs. We also summarize recent developments of Bayesian priors for calibrating computer models by field or experimental observations.  Finally, we review software packages for computer model emulation and calibration. 
\end{abstract}

\section{Prior distributions of statistical emulators for computer models with scalar-valued outputs}

\subsection{Parameters in Gaussian processes}  \label{model-description}

 Consider a simulator with input variables $\bm x =(x_1,\ldots,x_p) \in \mathcal{D} \subset \mathbb{R}^p$ and a scalar--valued function $f(\cdot) \in \mathbb{R}$, modeled by a Gaussian process 
 with mean and covariance functions
 \begin{align*}
   \mbox{E}[f(\bm x)] & =  \sum^q_{j=1} h_j(\bm x) \beta_j,  \\
   {\rm Cov}[f(\bm x), f(\bm x')] & = \sigma^2 c(\bm x, \bm x'; \bm \phi),
   \quad \bm x, \bm x' \in \mathcal{D},
 \end{align*}
that depend on the mean parameters $\bm \beta = (\beta_1,\ldots,\beta_q)^T$, variance $\sigma^2$ and correlation parameters 
$\bm \phi$.  
The terms $h_j(\bm x)$s are known mean basis functions and for every $\bm \phi$,
$c(\bm x, \bm x'; \bm \phi)$ is a correlation function on $\mathcal{D}$.
Then, for any design $\mathcal{S} = \{\bm x_1,\ldots,\bm x_n\}$
the joint distribution of $\bm f = (f(\bm x_1),\ldots,f( \bm x_n))^T$ 
is the multivariate normal distribution
 \begin{equation}
(\bm f \mid \bm \beta, \,\sigma^2, \, \bm \phi) \sim N (\mathbf H \bm \beta, \sigma^2   {\mathbf C}  ),\,
 \label{equ:GP}
 \end{equation}
where $\mathbf H =(\bm h^T(\bm x_1), \ldots,\bm h^T(\bm x_n))^T$
is the $n \times q$ matrix of mean basis functions, with $i^{\rm th}$ row
$\bm h(\bm x_i) = (h_1(\bm x_i),\ldots,h_q(\bm x_i))$, and
${\mathbf C}$ is the $n \times n$ matrix with $(i,j)$ entry $c(\bm x_i, \bm x_j; \bm \phi)$.
The mean function of the simulator is often set to be a constant, i.e. $q=1$ and $h(\bm x)=1$, 
although appropriate modeling of the mean can improve the predictive accuracy of a GP emulator for some functions \citep{Bayarri09}.  
 
 The covariance function in GP emulation is often assumed to be stationary due to computational considerations and lack of response replications.  
 Two classes of stationary covariance functions are commonly used:
 isotropic and separable covariance functions. 
 Isotropic covariance functions depend on the Euclidean distance between inputs, i.e. $\sigma^2 c(\bm x, \bm x'; \bm \phi) =\sigma^2 c(||\bm x - \bm x' ||; \bm \phi)$, which are often used to model spatially correlated data \citep{cressie1993statistics,gelfand2005spatial}. 
 In recent years, GP emulators with isotropic covariance functions have also been used to emulate potential energy surfaces, particle densities, and atomic forces of molecules for expensive simulations of molecules and materials \citep{rupp2012fast,bartok2013representing,brockherde2017bypassing,chmiela2017machine,li2022efficient,fang2022reliable}, where the inputs or descriptors in these applications may include inverse pair-distances of atomic positions, external potential functions, 
 or particle density functions, which are generally high-dimensional  \citep{deringer2021gaussian}. 
 Some commonly used isotropic correlation functions are given in Table \ref{tab:corr}, where $\phi$ is called a range parameter and
 $\alpha$ is called the smoothness or roughness parameter. 
 In practice, $\phi$ is assumed unknown while $\alpha$ is held fixed
 in commonly used packages for computer model emulation and calibration, but recent studies show the data also contains information for the roughness parameters, which will be discussed in Section \ref{packages}.
 Separable covariance functions are a product of the variance parameter and $p \geq 2$ stationary correlation functions, 
 where each factor models the correlation along one coordinate of the input vector. Separable covariance functions are often more sensible than isotropic covariance functions when the inputs have different natures and units, a common scenario in computer model emulation and calibration \citep{sacks1989design,bayarri2007framework,higdon2008computer}. 

 For any $\bm x=( x_{1}, \ldots,  x_{p})$ and $\bm x'=( x'_{1}, \ldots,  x'_{p})$ in $\mathcal{D}$, a separable covariance function is written as $\sigma^2 c(\bm x, \bm x'; \bm \phi)=\sigma^2 c_1( d_1;\phi_1) \times \ldots \times  c_p( d_p; \phi_p)$, 
 where $d_l=|x_l-x'_l|$ and $c_l(\cdot)$ is a stationary correlation function in $\mathbb{R}$ for the $l$th coordinate, 
In a separable covariance function, there is typically a single correlation parameter associated with each input variable, which controls the correlation length-scale between the outputs when this input variable changes. 
Input variables of different natures and units can then be accommodated by a separable covariance function.
Each $c_l(\cdot)$ in a separable covariance function is usually chosen from the same family of correlations, e.g., one of those in Table \ref{tab:corr}.
\begin{table}[t]
\caption{Popular choices of correlation functions. Here,
$\alpha$ is the roughness parameter often held fixed and $\phi$ is an unknown range parameter. The range of the parameters are given in the last two columns. $\Gamma (\cdot )$ is the gamma function and $\mathcal{K}
_{\alpha}(\cdot )$ is the modified Bessel function of the second kind of order $\alpha$.}
\label{tab:corrfun}
\centering
\begin{tabular}{cccc}
\hline
 Name       & ${c(d)}$     & $\alpha$ & $\phi$    \\      
 \hline
Power exponential  & $ \exp \left\{-\left(\frac{d}{\phi} \right)^{\alpha }\right\}$& $(0,2]$ & $(0,+\infty)$             \\    
Spherical          & $ \left\{ 1-\frac{3}{2} ( \frac{d}{\phi } ) + \frac{1}{2} ( \frac{d}{\phi } ) ^{3} \right\} \mathbf{1}_{[d/\phi \leq 1]}$  & / & $(0,+\infty)$ \\
Rational quadratic & $ \left\{ 1+ \left( \frac{ d }{\phi } \right) ^{2} \right\} ^{-\alpha }$ & $(0,+\infty )$ & $(0,+\infty )$ \\
Mat\'{e}rn         & $\frac{1}{2^{\alpha -1}\Gamma (\alpha )} ( \frac{\sqrt{2\alpha} d}{\phi } ) ^{\alpha } \mathcal{K}_{\alpha } ( \frac{\sqrt{2\alpha} d}{\phi } ) $ & $ (0,+\infty)$ & $(0,+\infty)$ \\
\hline
\end{tabular}
\label{tab:corr}
\end{table}
The separable power exponential and Mat{\'e}rn correlation functions are the most widely used, where the former is given by
\begin{equation}
c(\bm x, \bm x')=\exp\left( -\sum^p_{l=1} \Big(\frac{d_l}{\phi_l}\Big)^{\alpha_l}\right) .
\label{equ:prod_pow_exp}
\end{equation}
When $\alpha_l=2$ for all $l$, the correlation is called the Gaussian or squared exponential correlation function  \citep{rasmussen2006gaussian}. A GP with this correlation function is infinitely mean squared differentiable along all directions. 
When $\alpha_l=1$ for all $l$, the correlation is called the exponential correlation function, and a GP with this correlation function is not mean squared differentiable along any direction. 
It is worth noting that the GPs with power exponential correlation functions are also not mean squared differentiable for any 
$\alpha_l<2$. 

Unlike power exponential correlation functions,
Mat{\'e}rn correlation functions have closed-form expressions only when the smoothness parameter is a half--integer i.e. 
$\alpha_l = k_l +1 /2$ with $k_l \in \mathbb N_0$. When $\alpha_l=5/2$ for all $l$, for instance, the separable Mat{\'e}rn 
correlation function is given by
\begin{equation}
 c(\bm x, \bm x'; \bm \phi)= \prod^p_{l=1}\left(1+\frac{\sqrt{5}d_l}{ \phi_l}+\frac{5d^2_l}{3 \phi_l^2}\right)\exp\left(-\frac{\sqrt{5}d_l}{ \phi_l}\right).
\end{equation}
A good property of the Mat{\'e}rn covariance family is that the smoothness of the process is controlled by the smoothness 
parameters $\alpha_l$. 
A GP with a separable Mat\'ern correlation function 
is $\lfloor \alpha_l \rfloor$ times mean square differentiable along direction $l$, and thus one can control the degree of differentiability of the process by the choice of the smoothness parameters $\alpha_l$. 

{We have so far introduced a parametrization for the GP emulator for a deterministic computer model. 
The numerical solvers of some computers have non--negligible error and some computer model simulations can be intrinsically stochastic \citep{baker2022analyzing}.  Even for deterministic computer simulations with negligible numerical error, omitting some inert inputs can simplify the computation  \citep{gramacy2012cases,andrianakis2012effect,Gu2016PPGaSP}. In all these scenarios, given a set of inputs, the model output does not exactly match the computer simulation.  
To account for this uncertainty, a small noise term, sometimes called ``nugget" or ``jitter", is often included in the modeling, 
such that the observations follow the model
\[\tilde f(\bm x) = f(\bm x) + \epsilon,\]
where the noise follows $\epsilon\sim N(0,\sigma^2_0)$, with $\sigma^2_0 > 0$, and $\epsilon$ is independent of $f(\cdot)$.  
In this case, the joint distribution of the model output,
$\bm {\tilde f} = (\tilde f(\bm x_1),\ldots,\tilde f(\bm x_n))^T$, is
$(\bm {\tilde f} \mid \bm \beta, \,\sigma^2, \, \bm \phi) \sim N(\mathbf H \bm \beta, \sigma^2(\mathbf C + \eta \mathbf I_n))$, with $\eta = \sigma^2_0/\sigma^2$ 
and $\mathbf I_n$ is the $n \times n$ identity matrix}. 
To simplify notation, the correlation parameters in this case are augmented, i.e. $\phi_{p+1} = \eta$. 
Regardless of the absence or presence of a nugget, the GP  emulator with a separable covariance function  typically includes  
the parameters in the mean function $\bm \beta \in \mathbb{R}^q$, 
a variance parameter $\sigma^2 > 0$ and correlation parameters 
$\bm \phi$ in $\mathbb{R}_{+}^p$ or $\mathbb{R}_{+}^{p+1}$. 

In the next section, we discuss default prior distributions only for the parameters of GP emulators with separable covariance functions since prior distributions for isotropic covariance functions can be typically obtained from the former by restricting the number of range parameters to one.
Also, we will only describe prior distributions for the case when there is no nugget parameter since
these prior distributions can be easily extended to the case when there is a nugget parameter.

\subsection{Default prior distributions}  \label{reference-prior}

The specification of prior distributions for the parameters of Gaussian random fields is a challenging task.
First, subjective elicitation is often precluded by the lack of subjective 
information about the range parameters and the dependence of these parameters on the scale and nature of the input variables.
Second, commonly used naive priors often result in improper posterior distributions
(\cite{berger2001objective}, \cite{paulo2005default}). 
A theoretically attractive solution is the use of information--based default priors, the most prominent one being reference priors.
In a sense, these priors convey minimal information about the parameters relative to that conveyed by the likelihood, so the data becomes dominant for posterior inference.

Early priors for the parameters of isotropic models appeared in \cite{kitanidis1986} and \cite{handcock1993bayesian}, 
where partially conjugate priors for $(\bm \beta, \sigma^2)$ are proposed, but with little or no guidance for the prior specification of $\bm \phi$. 
The first systematic study of default priors for the parameters of GPs was undertaken by \cite{berger2001objective} for isotropic covariance models. Later, \cite{paulo2005default} studied reference and two variants of Jeffreys priors for GPs with separable covariance functions. Both articles assume the output is scalar-valued. All these default priors are partially conjugate for 
$(\bm \beta, \sigma^2)$ and have the general form 
\begin{equation}
\pi(\bm \beta, \sigma^2, \bm \phi) \propto \frac{\pi( \bm \phi)}
 { (\sigma^2)^a},
 \label{equ:prior_overall}
 \end{equation}
 where $\pi(\bm \phi)$ is a `marginal' prior for range parameters $\bm \phi$ and $a$ is a hyperparameter;
 $a=1$ is assumed in this work as it is the most widely used choice in practice. 
In \cite{berger2001objective}, the authors extensively discuss default priors and their respective posterior propriety for isotropic covariance models. It was shown that a certain reference prior always yields proper posterior distributions and inferences based on these priors display good frequentist properties. Although Jeffreys--rule priors also yield proper posteriors, inferences based on these do not display good frequentist properties.
Independent Jeffreys priors result in improper posterior distributions, so these priors are unsuitable to use for full Bayesian inference. 
On the other hand, \cite{paulo2005default} showed for separable covariance models that reference and both Jeffreys priors yield proper posteriors, and inferences based on reference and independence Jeffreys priors display similarly good frequentist properties. 
In this work, we restrict attention to reference priors or approximations thereof.


Without loss of generality, we assume the model does not contain a nugget parameter. 
The computation of reference priors requires the ordering of parameters according
to their inferential importance, which induces a factorization of the joint prior. 
For the models treated here, the covariance parameters are considered the most important and the mean parameters are less important, leading to the factorization
$\pi^{\rm R}(\bm \beta, \sigma^2, \bm \phi) =
\pi^{\rm R}(\bm \beta \mid \sigma^2, \bm \phi) \pi^{\rm R}(\sigma^2, \bm \phi)$.
The first factor is chosen as the Jeffreys--rule prior for $\bm \beta$ when
$(\sigma^2, \bm \phi)$ are assumed known, which for model (\ref{equ:GP}) is 
$\pi^{\rm R}(\bm \beta \mid \sigma^2, \bm \phi) \propto 1$. 
The second factor is also chosen as the Jeffreys--rule prior, but for the marginal model defined by the integrated likelihood of 
$(\sigma^2, \bm \phi)$
\begin{align}
p(\bm f \mid \sigma^2, \bm \phi ) & = \int_{\mathbb{R}^q} 
p(\bm f \mid  \bm \beta,\sigma^2,\bm \phi)
\pi^R(\bm \beta \mid \sigma^2, \bm \phi) d \bm \beta \nonumber \\
& \propto (\sigma^2)^{-\frac{n-q}{2}}
|{\mathbf C}|^{- \frac{1}{2}} |{\mathbf H^T { \mathbf C^{ - 1}}\mathbf H }|^{- \frac{1}{2}}  
\exp\left(-\frac{S^2}{2 \sigma^2}\right), 
\label{equ:marginal}
\end{align} 
where $\mathbf H$ is the $n \times q$ mean basis matrix and $S^2=\bm f^T \mathbf Q \bm f=(\bm f- \mathbf H \bm {\hat \beta} )^T \mathbf C ^{-1}(\bm f- \mathbf H \bm {\hat \beta} )$
with $\mathbf { Q} = { \mathbf C ^{ - 1}}\mathbf { P}$, $\mathbf { P} = \mathbf I_n - \mathbf H {\left({\mathbf H^T { \mathbf C ^{ - 1}}\mathbf H }\right)^{ - 1}}  \mathbf H^T {\mathbf C }^{ - 1}$ and $ \hat {\bm \beta}= {\left(\mathbf H^T { \mathbf C ^{ - 1}}\mathbf H \right)^{ - 1}}  \mathbf H^T {\mathbf C }^{ - 1} \bm f$ is the generalized least squares estimator of $\bm \beta$. 
The reference prior of the GP emulator with a separable covariance has the form (\ref{equ:prior_overall}) with the marginal prior for $\bm \phi \in \mathbb{R}_{+}^p$ given by 
\begin{equation}
\pi^{\rm R}(\bm \phi) \propto |\mathbf I^*(\bm \phi)|^{\frac{1}{2}}, 
\label{joint-ref-prior}
\end{equation}  
where $\mathbf I^*(\bm \phi)$, the Fisher information of the marginal model (\ref{equ:marginal}), is given by 
 \begin{equation}
 {\mathbf I^*}({\bm \phi}) = {\left( {\begin{array}{*{20}{c}}
   {n - q} & {\tr({\mathbf W_1})} & {\tr({\mathbf W_2})} & {...} & {\tr({\mathbf W_p})}  \\
   {} & {\tr(\mathbf W_1^2)} & {\tr({\mathbf W_1}{\mathbf W_2})} & {...} & {\tr({\mathbf W_1}{\mathbf W_p})}  \\
   {} & {} & {\tr(\mathbf W_2^2)} & {...} & {\tr({\mathbf W_2}{\mathbf W_p})}  \\
   {} & {} & {} &  \ddots  &  \vdots   \\
   {} & {} & {} & {} & {\tr(\mathbf W_p^2)}
 \end{array} } \right) },
 \label{equ:ref_prior_range}
 \end{equation}
with ${\mathbf W_l} = {\dot {{\mathbf C}}_l}\mathbf Q$ and ${\dot {{\mathbf C}} _l}$ is the partial derivative of the correlation matrix $\mathbf C$  with respect to the parameter $\phi_l$, for $l =1,\ldots,p$;
the latter have closed-form expressions for the frequently used covariance functions listed in 
Table \ref{tab:corr}.
{The reference prior 
(\ref{joint-ref-prior})--(\ref{equ:ref_prior_range}) was derived by \cite{berger2001objective} for models with $p=1$ and by \cite{paulo2005default} for models with $p \geq 2$.
\cite{ren2012objective,ren2013objective} provided reference priors for other orderings of the model parameters, and showed that some of these priors agree with (\ref{joint-ref-prior})--(\ref{equ:ref_prior_range}) and some do not. 
For models with a nugget, the expression for the reference prior agrees with that in (\ref{joint-ref-prior})--(\ref{equ:ref_prior_range}), after replacing everywhere $p$ with $p+1$ and $\mathbf C$ with $\mathbf C + \eta \mathbf I_n$ (recall that $\phi_{p+1} = \eta$). 
This generalization was studied in
\cite{de2007objective,kazianka2012objective,ren2012objective} for isotropic covariance models, and in \cite{Gu2018robustness} for separable covariance models. 
It is worth noting that the above expression for the reference prior holds for any correlation function, but the prior and posterior properties depend on the assumed correlation model}.




{
To use the reference prior (\ref{equ:prior_overall}) with the marginal prior for the range parameters (\ref{joint-ref-prior})  
for full Bayesian inference, we need to ensure the posterior distribution is proper.
This requires that the integral
\begin{equation*}
\int_{\mathbb{R}_{+}^p} p(\bm f \mid \bm \phi) \pi^R(\bm \phi) d \bm \phi ,
\end{equation*}
is finite, where 
\begin{equation}
p(\bm f \mid \bm \phi) \propto |{\mathbf C}|^{- \frac{1}{2}} 
|{\mathbf H^T { \mathbf C^{ - 1}}\mathbf H }|^{- \frac{1}{2}}  
(S^2)^{-\frac{(n-q)}{2}} ,
\label{equ:marginal-phi}
\end{equation}
is the marginal (integrated) likelihood of $\bm \phi$.
Determining the existence of the above integral requires knowledge of the asymptotic behavior of both 
$p(\bm f \mid \bm \phi)$ and $\pi^{\rm R}(\bm \phi)$ as $\bm \phi$ approaches the boundary of $\mathbb{R}_{+}^p$.
For isotropic covariance functions (i.e. the range parameter $\phi \in \mathbb  R_{+}$),  \cite{berger2001objective}
 put forward a set of technical but checkable assumptions that make this feasible from the Maclaurin expansion of the correlation function. 
The two most important assumptions are:
\begin{assumption}
For any $d \geq 0$, $c(d)= c^{0}(d/\phi)$, where $c^{0}(\cdot)$ is a continuous correlation function that satisfies 
$\displaystyle \lim_{u \to \infty }{c^{0}(u)}=0$.
\label{assumption:cont}
\end{assumption}
\begin{assumption}
As $\phi \to \infty$
\begin{equation}
\mathbf C = {\bf 1}_{n}{\bf 1}_{n}^T +\nu(\phi)(\mathbf{D} + o(1)),
\label{sigma-expansion}
\end{equation}
where ${\bf 1}_{n}$ is the vector of ones, $\nu(\phi) > 0$ is a continuous function satisfying 
$\displaystyle\lim_{\phi \rightarrow \infty}\nu(\phi) = 0$, and
$\mathbf{D}$ is a fixed $n \times n$ non--singular matrix.
\end{assumption}
Until recently it was thought that most isotropic correlation functions used in practice satisfy these assumptions.
Although this is certainly the case for Assumption 1, \cite{mure2021propriety} noticed that Assumption 2 does not hold
for some smooth correlation functions because for those $\mathbf{D}$ is singular. 
This is the case, for instance, for Mat\'ern models with $\alpha \geq 1$ and the power exponential model with 
$\alpha = 2$ (the latter was already noticed by \cite{paulo2005default}). 
Thus, the proof of propriety of the reference posterior in \cite{berger2001objective} is valid only for 
non--smooth families of isotropic correlation functions.
For all covariances considered in \cite{berger2001objective}, including both smooth and non--smooth covariances,  posterior propriety still holds in general since \cite{mure2021propriety} provided a proof for the case of smooth families of isotropic correlation functions that does not require $\mathbf{D}$ to be non--singular.
Assumption 2, or a slight variation of it, was used in virtually all works that provided reference posterior propriety for a variety of isotropic and separable models 
\citep{paulo2005default,kazianka2012objective,ren2012objective,ren2013objective,Gu2018robustness}.
As a result, the proofs of propriety of reference posteriors in these works are also incomplete 
and in need of completion for smooth families of separable covariance functions, and the same holds when a nugget parameter is included; 
see \cite{mure2021propriety} and \cite{de2022approximate} for further discussion. 

For the GP emulator context, however, the range parameters are rarely estimated in a full Bayesian way, due to the large computational cost of carrying out Markov chain Monte Carlo (MCMC) sampling. Instead, the maximum likelihood estimator or the marginal posterior mode estimator of the range parameters is often used. The estimated range parameters are then plugged into the predictive distributions (see Section \ref{subsec:param_est} for more discussions). Therefore, one does not need to worry about the posterior propriety of $\bm \phi$ in a GP emulator, as $\bm \phi$ is typically estimated and plugged into the predictive distribution, which is a proper distribution. 
}

{{\bf Approximations}.
Evaluation of the reference prior (\ref{joint-ref-prior})--(\ref{equ:ref_prior_range}) is
computationally demanding, even for isotropic models and when the sample size is moderate. 
The computational problem is threefold. First, evaluating the entries of $\mathbf I^{*}(\bm \phi)$ involves computing terms such as ${\mathbf C}^{-1}\bm f$ and ${\mathbf C^{-1}}\mathbf H$, 
which requires $\mathcal O(n^3)$ operations, since these involve the Cholesky decomposition of $\mathbf C$. 
Second,  the derivatives of the reference prior with respect to the range parameters are hard to compute. 
When the marginal posterior mode estimator is used for estimating the range parameters, however, 
 derivatives of the marginal posterior distribution are often required in numerical optimization \citep{nocedal1980updating}. Numerical derivatives, on the other hand,  are more time-consuming and less stable than derivatives with closed-form expressions.   
Third, computation of the logarithm of the reference prior is more unstable than computation of the logarithm of the marginal likelihood, when the covariance matrix $\mathbf C$ is nearly singular.


To ameliorate these computational challenges, \cite{gu2018jointly} proposed a computational scalable prior $\pi^{\rm JR}(\bm \phi)$, dubbed jointly robust (JR), aiming at matching the behavior of the reference prior $\pi^{\rm R}(\bm \phi)$ when $\bm \phi$ 
approaches the boundary of the parameter space. 
Denote the inverse range parameters as $\bm {\tilde \phi}=(\tilde \phi_1,\ldots,\tilde \phi_p)$, with $\tilde \phi_l=1/\phi_l$, 
for $l=1,\ldots,p$. The jointly robust prior is given by 
$\pi^{\rm JR}(\bm \beta, \sigma^2,\bm {\tilde \phi}) \propto \pi^{\rm JR}(\bm {\tilde \phi})/\sigma^2 $, with 
\begin{equation}
\pi^{\rm JR}(\bm {\tilde \phi})=w_0 \left(\sum^p_{l=1}w_l  {\tilde \phi}_l \right)^{b_1}\exp\left(-b_2\sum^p_{l=1}w_l  {\tilde \phi}_l \right),
\label{equ:JR_prior}
\end{equation}
where $b_1 > -(p+1)$, $b_2 > 0$, $w_l > 0$ are hyperparameters and $w_0=\frac{p!b_2^{b_1+p+1}\prod^p_{l=1} w_l}{\Gamma(b_1+p+1)}$ 
is the normalizing constant;  \citep{gu2018jointly} provides guidelines for choosing these hyperparameters.
When $p=1$, this prior becomes a Gamma prior for the inverse range parameter, which has desirable convergence property for isotropic kernels \citep{van2009adaptive}. When $p \geq 2$, the marginal posterior mode estimation with this prior avoids the unstable estimation of the range parameters, similar to the reference prior (see more discussions in Section \ref{subsec:param_est}). 
Another benefit of this prior is that it can shrink the size of the inverse range parameter when used in the marginal posterior mode estimation,  useful for identifying the `inert inputs' i.e. inputs with little or no effect on the computer model \citep{linkletter2006variable}. 

More recently, for isotropic models without a nugget, \cite{de2022approximate} developed an easy to compute approximation of the reference prior $\pi^{\rm R}(\phi)$ based on an approximation to the integrated likelihood of
the range parameters (\ref{equ:marginal}) derived from the spectral approximation of stationary
random fields. 
The computation of this approximate reference prior relies on the spectral density function of the 
random field, which does not involve the inversion of large or ill--conditioned matrices and has a
matrix--free expression for models with constant mean; 
see \citep{de2022approximate} for Bayesian and frequentist properties of inferences based on this approximate reference prior.
}


   

\subsection{Parameter estimation and predictive distribution}  \label{subsec:param_est}

In building a GP emulator for approximating expensive computer models, full posterior inference typically requires  MCMC samples, which can be computationally expensive, as each evaluation of the posterior distribution takes $\mathcal O(n^3)$ operations, with $n$ being the sample size. As an alternative, one can use a point estimator of the correlation parameters. Next, we discuss a few ways to estimate the range parameters and illustrate how reference prior distributions can make the estimation more robust.   

Maximum likelihood estimators (MLE) and maximum marginal likelihood estimators (MMLE) are often used to estimate correlation parameters, where the latter is obtained by maximizing w.r.t. $\bm \phi$ the
marginal likelihood (\ref{equ:marginal-phi}). 
However, these methods sometimes select extreme values of the correlation parameters, so that the estimated covariance matrix becomes $\hat {\mathbf C} \approx \mathbf  I_n$ or $\hat {\mathbf C} \approx {\mathbf 1_n \mathbf 1_n^T}$, which results in unstable estimation and prediction. 

A way to avoid the two extreme scenarios is to estimate the range   parameters using maximum marginal posterior estimation (MMPE),  
${\bm {\hat \phi} }=\argmax_{\bm \phi} \log(p( \bm f \mid \bm \phi )  \pi(\bm \phi))$, 
where $p( \bm f \mid \bm \phi )$ is the marginal likelihood (\ref{equ:marginal-phi}) and $\pi(\bm \phi)$ is the prior distribution for the correlation parameters.
It was shown in \cite{Gu2018robustness} that, when the covariance function is parameterized  either in terms of the range parameters (Table \ref{tab:corr}) or the logarithm of the inverse range parameters,
the MMPE with a 
 reference prior (\ref{joint-ref-prior}) of some frequently used separable correlation functions is robust in the sense that the aforementioned extreme scenarios
In practice, an approximate prior, such as the JR prior in Equation (\ref{equ:JR_prior}), can be used for speeding up the numerical optimization of finding the marginal posterior mode to estimate range parameters, as closed-form expressions of the derivative of 
the marginal posterior distribution with respect to the range parameters are often available. The cost of computing the derivative 
of the marginal posterior with the reference prior is too high, and thus numerical derivatives are often used instead. 

  


Let $\bm x^*$ be a test input.  
We can carry out predictive inference about $f(\bm x^*)$ based on the predictive distribution $(f(\bm x^*) \mid   \bm f,  \bm {\hat \phi})$,
obtained by integrating out the mean and variance parameters w.r.t  their reference prior in (\ref{equ:prior_overall}). 
This results in a non-central Student's t distribution with $n-q$ degrees of freedom, 
\begin{align}
( f(\bm x^*) \mid   \bm f,  \bm {\hat \phi} )  & \sim  
t( \hat f({\bm x}^{*}),\hat{\sigma}^2   c^{**}, n-q), 
\label{equ:pred_dist_scalar}
\end{align}
 where 
\begin{align}
    \hat f ({\bm x}^{*}) &= { \bm h({\bm x}^{*})} \hat{\bm{\beta}}+\bm{c}^T(\bm{x}^*){\hat{\mathbf C}}^{-1}\left(\bm f-\mathbf H\hat{\bm{\beta}} \right), \label{equ:gppredmean}\\
        \hat{\sigma}^2 &= {(\bm{f}-\mathbf H \hat{\bm{\beta}})}^{T}{\hat{\mathbf C}}^{-1}({\bm{f}}-\mathbf H\hat{\bm{\beta}})/(n-q), \label{equ:sigma_j_hat} \\
    c^{**} &= c^{*} +  {\bm  h}^*(\bm x^*)^T \left(\mathbf H^T{\hat{\mathbf C}}^{-1}\mathbf H \right)^{-1} {\bm  h}^*(\bm x^*) ,  \label{equ:gppredcorrelation}
\end{align}
with $\bm{c}(\bm{x}^*) = (c(\bm{x}^*,{\bm{x}}_1 ), \ldots, c(\bm{x}^*,{\bm{x}}_n ))^T$, $c^*=c({\bm x^{*}}, {\bm x^{*}})-{ \bm{c}^T(\bm{x}^*){ \hat{\mathbf C}}^{-1}\bm{c}(\bm{x}^*)}$,   
${\bm h}^*(\bm x^*)={{\bm h(\bm x^{*})}}-\mathbf H^T {\hat{\mathbf C}}^{-1}\bm{c}(\bm{x}^*)$, and ${\hat{\bm \beta}}=\left( \mathbf H^T {\hat{\mathbf C}}^{-1} \mathbf H \right)^{-1}\mathbf H^T{\hat{\mathbf C}}^{-1}\bm{f}$ being the estimated generalized least squared estimator. 
The predictive mean or median $\hat f ({\bm x}^{*})$ is often used for predicting $f(\mathbf x^*)$, while the uncertainty in the prediction can be obtained from appropriate quantiles of the predictive distribution, which are easily available from (\ref{equ:pred_dist_scalar}).  







\section{Prior distributions of statistical emulators for computer models with vector-valued outputs}

Some computer models have not one but many outputs from massive spatial or spatio-temporal coordinates associated with each input vector. 
For instance, the TITAN2D simulator \citep{pitman2003computing,patra2005parallel} produces flow properties, such as the height and velocity of the pyroclastic flow, in a volcanic eruption at a large number of space--time coordinates, for a given set of input parameters, such as volume of the chamber, initial eruption angles, basal and internal friction angles. 
The updated TITAN2D  model takes roughly 8 minutes to run per set of inputs in a desktop \citep{simakov2019modernizing}, 
while a large number of runs, of the order of $10^5$--$10^6$, are often required for hazard quantification. 
Hence, a fast statistical surrogate model is needed to emulate the many outputs at a massive number of coordinates. 
 
 
 For a simulation run at a physical input $\bm x$, consider the
 computer model output associated to this input at $k$ grid points $\bm s_1,\ldots,\bm s_k$,  denoted as 
 $\bm f(\bm x)=(f(\bm s_1, \bm x),\ldots,f(\bm s_k, \bm x))$, and for $j=1,\ldots,k$, we use
 $f_j(\bm x) = f(\bm s_j, \bm x)$ as a shorthand. 
 For simplicity, assume the computer model produces output at the same set of grid points for each simulation input, 
 so the output for $n$ simulation runs with inputs $\bm x_1,\ldots,\bm x_n$ is the $k\times n$ matrix 
 $\mathbf F=[\bm f(\bm x_1),\ldots,\bm f(\bm x_n)]$. 
 Many choices have been explored for modeling the output vector $\mathbf F$. For instance, the 
 separable model was introduced in \cite{conti2010bayesian}, which assumes that  
 \begin{align}
 \left(\mathbf F \mid \mathbf B, \bm \Sigma, \bm \phi \right) \sim  N_{k,n}\left( \mathbf B^T \mathbf H^T, \bm \Sigma, \mathbf C \right)
 \label{equ:separable_model}
 \end{align}
 where $N_{k,n}$ denotes the matrix normal distribution with dimensions $k\times n$, $\mathbf H$ is a $n\times q$ matrix of mean basis,  $\mathbf B=[\bm \beta_1,\ldots,\bm \beta_k]$ is an $q\times k$ matrix of trend parameters, with 
 $\bm \beta_j=(\beta_{j1},\ldots,\beta_{jq})^T$ being the $q$-vector of trend  parameters at coordinate $j$, 
 $\bm \Sigma$ is the $k\times k$ covariance matrix of the output across grid points and $\mathbf C$ is the $n\times n$ correlation matrix of output between the physical inputs, often modeled by a correlation function with range parameters $\bm \phi$. 
 This model assumes that for inputs $\bm x, \bm x'$ and grid points $\bm s_j, \bm s_{j'}$, 
 Cov$(f(\bm s_j, \bm x), f(\bm s_{j'}, \bm x')) = \Sigma_{j,j'} c(\bm x,\bm x')$, where $\Sigma_{j,j'}$ is the 
 $(j,j')$th entry is the covariance matrix $\bm \Sigma$ between outputs at grid points $\bm s_j$ and $\bm s_{j'}$, 
 for $j,j'=1,\ldots,k$, and $c(\cdot,\cdot)$ is a separable correlation function as the ones discussed in 
 Section \ref{model-description}. Denote the transformed range parameters $\bm r=(r_1,...,r_p)^T$ with $r_l=1/\phi^2_l$ for $l=1,...,p$,  \cite{conti2010bayesian} proposed the  prior
 \begin{equation}
 \pi(\mathbf B, \bm \Sigma, \bm r)\propto \prod^p_{l=1} \left(1+ r^2_l\right)^{-1} |\bm \Sigma|^{-(k+1)/2}, 
 \end{equation}
 where $(\mathbf B, \bm \Sigma)$ can be integrated out, and the transformed range parameters $\bm r$ are estimated by the MMPE with the parametrization of $\bm r$ in both the marginal likelihood and prior.

Computing the likelihood of the separable model in (\ref{equ:separable_model}) takes  $\mathcal O(n^3)+\mathcal O(k^3)$ operations, due to inversion of the covariance matrices, which can be infeasible when either the number of grid points or the number simulation runs are large. 
 \cite{Gu2016PPGaSP} introduced the parallel partial Gaussian process (PP--GP) to overcome the large computational cost 
 when the number of grid points, $k$, is large. The output in the PP--GP model at the $j$th grid point,  
 $\bm f_j = (f_j(\bm x_1),\ldots,f_j(\bm x_n))$, is assumed to have different means and variances, but 
 the correlation is shared across the grid points. 
 That is, 
 $(\bm f_j\mid \bm \beta_j, \sigma^2_j, \mathbf C) \sim N(\mathbf H \bm \beta_j, \sigma^2_j \mathbf C) $, where $\bm \beta_j$ is a $q$-vector of the trend parameters, $\sigma^2_j$ is a scalar value of the variance parameter and $\mathbf C$ is the $n\times n$ correlation shared across the grid. Here the $n\times q$ mean basis matrix $\mathbf H$ could be different across grid points, 
 but for simplicity we assume they are the same. 
 Hence, the correlation between different grid points is assumed to be zero for computational purposes. 
 Interestingly, the predictive mean of PP--GP under this independent assumption is generally the same as that of the more general 
 separable model, as shown in Theorem 6.1 in  \cite{Gu2016PPGaSP}. 
 
 Following the procedure discussed in Section \ref{reference-prior},
 the reference prior can be extended to the PP--GP emulator for vector--valued output, resulting in
  \begin{equation}
 \pi(\bm \beta_1,\ldots, \bm \beta_k,\sigma^2_1,\ldots,\sigma^2_k,\bm \phi) \propto  
 \frac{|\mathbf I^*(\bm \phi)|^{1/2}}{\prod^k_{j=1}\sigma^2_j},
 \end{equation}
where $\mathbf I^*(\bm \phi)$ is obtained from the Fisher information matrix similar to that in (\ref{equ:ref_prior_range}).  
The range parameter $\bm \phi$ can be estimated by mode estimation, such as the MLE, MMLE, or MMPE,  and these are implemented 
in the {\tt RobustGaSP} package \citep{gu2018robustgasp}. 
Integrating out the mean and variance parameters, and plugging in the estimated range parameters, the predictive distribution of a test input $\bm x^*$  follows $( f_j(\bm x^*) \mid   \mathbf F,  \bm {\hat \phi} ) \sim  t( \hat f_j({\bm x}^{*}),\hat{\sigma}^2_j  c^{**},n - q)$, where $\hat f_j({\bm x}^{*})$ and $\hat{\sigma}^2_j$ are obtained by replacing $\bm f$ with 
$\bm f_j$ in (\ref{equ:gppredmean}) and (\ref{equ:sigma_j_hat}), respectively. 

 Computing the predictive mean in PP--GP model requires $\mathcal O(nk)+\mathcal O(n^3)$ operations, and computing the predictive variance requires an additional $\mathcal O(n^2k)$ operations. The computational complexity of fitting the PP-GP is smaller than building independent GP emulators for each grid point, which generally takes $\mathcal O(n^3k)$ operations due to the required Cholesky decomposition correlation matrices at $k$ grid points. Fitting the separable model in (\ref{equ:separable_model}) requires computing the Cholesky decomposition of covariance matrix across grid points, at $\mathcal O(k^3)$ operations, which can be expensive if the number of grid points is large. The computational advantage of the PP-GP makes it suitable for 
 applications with massive grid points, such as emulating geophysical flow simulation, such as the TITAN2D model \citep{Gu2016PPGaSP},  ocean circulation models for hurricane surges \citep{plumlee2021high}, landslide simulation \citep{zhao2021emulator},  and continuum-mechanical models of ground deformation of the Earth's surface \citep{anderson2024computationally}.

 \begin{figure}[t]
\centering
  \begin{tabular}{cc}
    \includegraphics[scale=1]{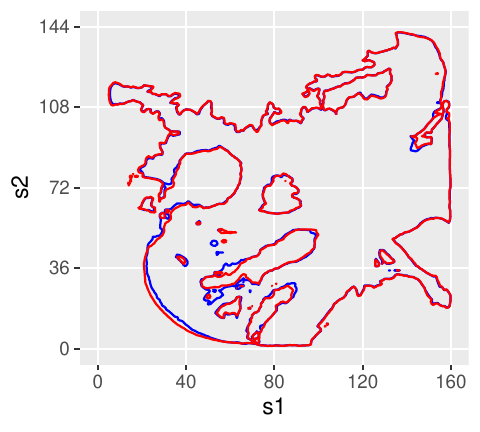}
        \includegraphics[scale=1]{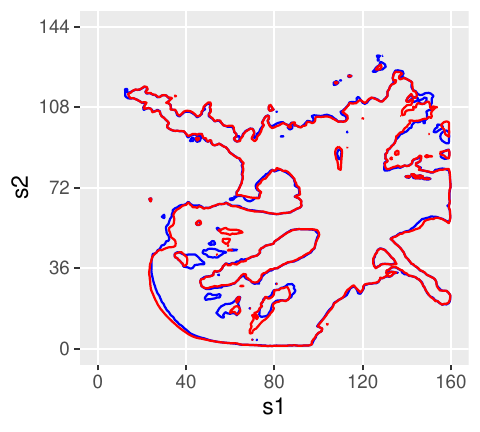}
        \vspace{-.1in}
  \end{tabular}
   \caption{One-meter contour of predictions (blue curves) and truth (red curves) of the first two held-out test inputs from TITAN2D by the PP-GP emulator. Here the inputs of the TITAN2D are chamber volume, initial eruption angle, basal friction angle and internal friction angle. The test inputs of the left and right panels are $\bm x=(10^{7.4}, 4.4, 11.0,21.9)^T$ and $\bm x=(10^{7.3}, 3.6,11.4, 22.9)^T$, respectively.  } 
\label{fig:titan2d}
\end{figure}

As an illustration, Figure \ref{fig:titan2d} gives the one-meter contour of the maximum pyroclastic flow heights of TITAN2D 
(maximum over time) at two held-out inputs, where the predictions are computed by the PP--GP emulator from the {\tt RobustGaSP} package \citep{gu2018robustgasp}. The area inside the contour has flow heights larger than one meter. Here for each input, TITAN2D produces the maximum flow height at $144\times 160$ grids, after maximizing the flow height over time in each simulation.  
The PP--GP emulator with $n=50$ runs already gives reasonably good predictions. It is worth mentioning that less than 20 seconds 
were required in a desktop for fitting the PP--GP model and making predictions on hundreds of test inputs. 
 

 
 
 



\section{Prior distributions and estimation for Bayesian model calibration} 

Computer models often contain unobserved parameters. Field data or experiments can be used to estimate parameters from the computer models, a process often termed computer model calibration or data inversion. In \cite{anderson2019magma}, for instance, observations from Interferometric Synthetic Aperture Radar (InSAR) and continuous Global Navigation Satellite System (GNSS) are used for calibrating  a continuum-mechanical model with unobserved calibration parameters, such as the centoid  of the magma chamber, magma density and pressure change rates for K\={\i}lauea Volcano in Hawaii, which experienced one of the largest eruptions that destroyed hundreds of
homes in 2018. 

 Denote the field observations $y(\bm x)$ at an observable input $\bm x$. When the computer model can reliably represent reality, one may model observations by $y(\bm x)=f(\bm x, \bm \theta)+\epsilon$, where $f(\bm x, \bm \theta)$ is a computer model at observed input $\bm x$ and unobserved calibration parameters $\bm \theta$, and 
$\epsilon\sim N(0,\sigma^2_0)$ denotes the Gaussian noise with variance $\sigma^2_0$. Here, the calibration parameters $\bm \theta$  have physical meanings, and the domain of these parameters is often compact.

There may be a discrepancy between the computer model and reality, leading to the model
\begin{equation}
y(\bm x)=f(\bm x, \bm \theta)+\delta (\bm x)+ \epsilon, 
\label{equ:calibration}
\end{equation}
where $\delta(\cdot)$ denotes a discrepancy function. In \cite{kennedy2001bayesian}, $\delta(\cdot)$  was modeled by a GP, and this approach was followed widely in later studies (see e.g.  \citep{bayarri2007framework,bayarri2007computer,higdon2008computer,paulo2012calibration}). 
Predictions by combining both the computer model and reality in (\ref{equ:calibration}) are often better than 
using either a computer model or a GP model alone.

Suppose, for simplicity, that $\delta (\cdot)$ has zero mean vector and covariance matrix $\sigma^2\mathbf C$, with $\bm \phi$ the range parameters in the covariance matrix $\mathbf C$. Assume we have $n$ field observations $\bm y=(y(\bm x_1),..., y(\bm x_n))^T$ and computer model outputs $\bm f(\bm \theta)=( f(\mathbf x_1,\bm \theta),...,f(\mathbf x_n,\bm \theta))^T$.  
After integrating out $\delta(\cdot)$ in (\ref{equ:calibration}), the marginal likelihood becomes a multivariate normal distribution
\[ (\bm y \mid \bm \theta, \sigma^2, \bm \phi) \sim N(\bm f(\bm \theta), \sigma^2\mathbf C +\sigma^2_0 \mathbf I_n ),  \]
where $\bm \theta$ become the parameters in the mean vector of a multivariate normal distribution. 
After reparameterizing the nugget parameter as $\eta=\sigma^2_0/\sigma^2$, the prior for the calibration model is given by
\[\pi(\bm \theta, \sigma^2,\bm \phi,\eta )\propto \frac{\pi(\bm \theta)\pi(\bm \phi,\eta)}{\sigma^2}.\]
Here $\pi(\bm \theta)$ is the prior distribution of the calibration parameters, chosen by the user, and a uniform prior for the calibration parameters is often used in practice. 
The prior $\pi(\bm \phi,\eta)$ is the prior for the range and nugget parameters, which is often set in the same way as the priors 
discussed in Section \ref{reference-prior}.
MCMC is often used to estimate the parameters in Bayesian model calibration, as the uncertainty of the calibration parameters can be quantified by posterior samples, which are often required in applications.

It was noticed that the estimated computer model in (\ref{equ:calibration}) can be far away from the reality  \citep{arendt2012quantification,tuo2015efficient}, in terms of some regularly used loss function, such as $L_2$ loss, due to the inclusion of a GP model as the discrepancy. The identifiability issue is often more pronounced when the correlation (i.e. entries in $\mathbf C$) is estimated to be large, as the main variability of the field data can be explained by the discrepancy function instead of the computer model.  In \cite{gu2018jointly}, the mode calibration with the JR prior was found to reduce identifiability issues compared with the use of the reference prior. In \cite{tuo2015efficient,wong2017frequentist}, two-step frequentist estimators were introduced for obtaining the ``$L_2$-minimizer" of calibration parameters, defined to be the ones that minimize the $L_2$ distance between the computer model and reality. Other than these two-step approaches, the scaled GP (S-GP )prior of the discrepancy was constructed in \cite{gu2018sgasp} for jointly using computer simulations and discrepancy models in model calibration. 
The S-GP prior places more prior mass of the random $L_2$ distance of the discrepancy function near the origin to alleviate the identifiability issue, whereas the variance parameter of the discrepancy and all other parameters are estimated by the data.  A scalable computational approach was also introduced in \cite{gu2018sgasp} to avoid sampling the discrepancy function modeled by the S-GP prior each time, and the convergence properties are studied in  \cite{gu2022theoretical}. 

\section{Computation and software packages}  \label{packages}

Mode estimators, such as MLE, MMLE and MMPE are often used in estimating parameters in GP emulators, because of their computational efficiency. 
A few software packages provide MLE of parameters and predictive distribution for emulating scalar--valued computer model, such as {\tt DiceKriging} \citep{roustant2012dicekriging} and {\tt GPfit} \citep{macdonald2015gpfit} and {\tt RobustGaSP} \citep{gu2018robustgasp}. The {\tt RobustGaSP} package allows users to emulate both scalar--valued and vector--valued computer models, and users can choose to use MLE, MMLE or MMPE for estimating the parameters. Both reference prior and JR priors are implemented in {\tt RobustGaSP} package. The default choice is to use the MMPE with the JR prior, as closed-from derivatives are available for numerical methods, such as low-storage quasi-Newton optimization methods \citep{nocedal1980updating} for finding the marginal posterior mode in estimating the range parameters. 
Other functionalities of {\tt RobustGaSP} package include estimating the ``inert input", defined as the input that barely affect the output of the computer simulations, and the periodic folding approach \citep{spiller2014automating} for defining the correlation function for a circular input, such as an angle input. All these packages implement routinely used correlation functions, 
such as power exponential correlation and Mat{\'e}rn correlation functions listed in Table \ref{tab:corrfun}.

A few packages implement computer model calibration in R, including {\tt BACCO} \citep{hankin2005introducing}, {\tt Save} \citep{palomo2015save},  {\tt Calico} \citep{carmassi2018calico} and {\tt RobustCalibration} \citep{gu2023robustcalibration}. Some computer models are computationally expensive, and emulators are required for accelerating computation in model calibration.  Both scalar-valued and vector-valued GP emulators by PP-GP, for instance, can be called in {\tt RobustCalibration}. MCMC is often used for computer model calibration in these packages. This is because the uncertainty of calibration parameters, required in most studies, can be obtained from MCMC samples.  
The computational scalability in model calibration is a bottleneck for many problems, and more advanced statistical learning approaches are needed to reduce the computational cost in computer model calibration. 

All of the above packages assume known smoothness or roughness parameters.
This is due partly to computational challenges when attempting to estimate all covariance parameters and partly to perceptions in the literature about the data having little or no information about the smoothness or roughness parameters $\alpha_l$. 
In the context of fitting the isotropic Mat\'ern covariance function to spatial data, \cite{de-han2022} showed that the information the data have about the smoothness parameter can be substantial, depending on the design and true model, and 
\cite{han-de2024approximate} derived an approximate reference prior for $\alpha$ in a simplified model.
It is to be determined whether accounting for the uncertainty about smoothness or roughness parameters could
improve the inference for expensive computer simulations.



\medskip

{\bf Acknowledgments}. 
 Mengyang Gu acknowledges the support from the U.S. National Science Foundation under award No. OAC-2411043. The work of Victor De Oliveira was supported by the U.S. National Science Foundation grant under Award No. DMS–2113375.

	\bibliographystyle{apalike}
	\bibliography{bibtex_example}

\end{document}